\documentclass{raa}            

\usepackage{graphicx,times}             
\usepackage{natbib}
\usepackage{amssymb,amsmath}
\bibpunct{}{}{;}{a}{}{,}
\usepackage[a4paper=true, pagebackref=true, bookmarks=false]{hyperref}
\hypersetup{colorlinks = true, linkcolor = green, anchorcolor = red, citecolor = blue, filecolor = red, pagecolor = red, urlcolor = red}

\begin{document}

   \title{Effects of dynamo magnetic fields on observational properties of Accreting Millisecond X-ray Pulsars
}

   \volnopage{Vol.0 (20xx) No.0, 000--000}      
   \setcounter{page}{1}          

   \author{D.~B, Zeleke
      \inst{1}
   \and S.~B, Tessema
      \inst{1}
   \and S.~H, Negu
      \inst{1}
   }

   \institute{$^1$Ehtiopian Space Science and Technology Institute (ESSTI),
              Entoto Observatory and Research Center (EORC), Astronomy and Astrophysics Research and Development Department,
              P.~O.~Box 33679 Addis Ababa, Ethiopia; {\it dugasa32@gmail.com}\\
                                 {\it tessemabelay@gmail.com}\\ {\it seblu1557@gmail.com}\\
\vs\no
   {\small Received~~20xx month day; accepted~~20xx~~month day}}
\abstract{In this paper, we have investigated the accreting millisecond X-ray pulsars, which are rapidly rotating neutron
stars in low-mass X-ray binaries. These systems show coherent X-ray pulsations that arise when the accretion flow 
is magnetically channeled to the stellar surface.  
Here, we have developed the  fundamental equations  for an accretion disc around accreting millisecond X-ray pulsars 
in the presence of a dynamo generated magnetic fields in the inner  part of the disc and we have also formulated 
the numerical method for the structure equations in the inner region of the disc and the highest accretion rate is 
enough to make the inner region of the disc which is overpowered by radiation pressure and electron scattering. 
Finally, we have examined our results with the effects of dynamo magnetic fields on accreting millisecond X-ray pulsars.
\keywords{Accretion disc, neutron stars, pulsars, millisecond X-ray pulsars }
}

   \authorrunning{D.~B. Zeleke et al. }            
   \titlerunning{Accretion disc: accreting millisecond X-ray pulsars }  

   \maketitle

%
%
\section{Introduction}           
\label{sect:intro}
Low-mass X-ray binaries (LMXBs) consists of  accreting  millisecond X-ray pulsars (AMXPs) from a low
mass evolved star (a degenerate dwarf star), which has rapidly rotating neutron stars (NSs). Hence, AMXPs 
are distinguished from the group of ordinary rotation-powered pulsars by their small spin periods \cite{Becker+2001}.
In these systems, the accreting matter may spin up the NS. Here, one of the possible endpoints of the evolution 
of a LMXB is expected to be  millisecond pulsar \cite{Strohmayer+2001discovery}.
Most of the  LMXB does not show coherent pulsation in their light curves due to that they are still
under debate; it could be due to the alignment of the  effect of the magnetic field on NS 
with its rotational axis. The only subclass of LMXB in which coherent pulsations have been observed is that the AMXPs.

A more observationally inclined review of accreting a millisecond pulsar is given by \cite{Wijnands+etal+2005chandra}.  
It is a transient system in which the outburst stage associated with the matter falling onto the NS surface and spin up its 
period in the order of a millisecond. Also, the first real AMXPs was studied by \cite{Wijnands+1998millisecond},
in which the spin frequencies range from 182 up to 599 Hz \cite{Falanga+2013accreting}.
Among these, the fastest accreting millisecond pulsar is IGR J 0029 + 5934 with the period of just 1.67ms \cite{Shaw+2005discovery};  \citep{Falanga+2005integral}.
It shows pulse frequency variations. These observations are very important for the understanding 
of the evolution of the NSs in LMXBs \cite{Poutanen+2006accretion}.

Here, in AMXPs, we have considered that the rapidly rotating NS has a weak magnetic dipole moments $\sim 10^{15} \rm{Tm^ 3}$ in the
inner region than the ordinary X-ray pulsars. In these systems, the accretion disc will be extended near to the NS 
and the temperature becomes more in which their opacity can be overpowered by radiation pressure and electron scattering \cite{Lasota+2016black}.
The magnetic fields are important for the transportation of angular momentum in these systems .
As it was studied by \cite{Tessema+2010structure} the region of the accretion disc which is located 
in the inner part of corotation radius supply spin-up torque on the NS while the outer part of the accretion disc 
brakes the NS.The resultant torque is investigated by the inner region of the disc position, which is displaced inwards as the
accretion rate increases.\\

AMXPs become important in many areas of astrophysical research. Here, It shows a very high average mass transfer 
rate $\rm{\dot{M}} = 10^{14}\rm{ kgs^{-1}}$ in the inner region and exhibit persistent X-ray pulsations with less than 10ms and  weak
magnetic fields.  Many authors have been studied accretion disc  in different models for
example \cite{Shakura+1973black};  \cite{Ghosh+1979accretion}. Hence, they didn't address an accretion disc in these systems 
particularly in the inner region of the disc. However,  \cite{Tessema+2010structure} were tried to study accretion disc
around magnetized stars using  pure models of magnetohydrodynamics (MHDs), but the present study will focus on 
the accretion disc in AMXPs in the inner region of the disc using analytical and numerical solutions.

In this study, we develop the fundamental equations of an accretion disc for
a dynamo generated an accretion disc around AMXPs, particularly we investigate the solution of 
these equations in the inner part of the disc using surface density, temperature, and radial 
velocity as a function of the radius  and finally, we present the numerical solution for the structure equations in
the inner region of the disc.\\
This paper is organized as follows:
In section (2) we investigate the fundamental equations of an accretion disc and we present the numerical method for
structure equation in the inner region of the disc. Hence, their results and discussion were incorporated in section (3) 
and finally, we summarize our results in section (4).

\section{Fundamental equations of accretion disc}
\subsection{Basic Assumptions}
In this study, we consider the accretion disc around  AMXPs with NS of mass $1.4\rm{M_{\odot}}$,  radius $10\rm{ km}$ 
and magnetic dipole moment $10^{15} \rm{Tm^3}$. 

Here, we have considered that the scale height of the disc, $H$ is much smaller than the radial extension of the disc, $R$ 
\cite{Shakura+1973black}.  
The gas in the disc rotates at Keplerian velocity and the orbital kinetic energy is transformed into radiation
by viscosity of an  accretion disc, $v$, while the angular momentum is transported outward.
\begin{equation}
v=\alpha_{ss}c_s H. \label{eqn:1}
\end{equation}
where $\alpha_{ss}$ $\sim 10^{-2}$ is  turbulence stress of the disc that describes the transport of angular 
momentum and that numerical simulations
 suggested by \citep{Hawley+1995local} and $c_s$ is the speed of sound in the gas.
 
 \subsection{Conservation of mass}
The law of conservation of mass or principle of mass conservation states that for any system closed to all transfers 
of matter and energy, the mass of the system must remain constant over time, as the system's mass cannot change, so the quantity
can not be added nor removed. Hence, the quantity of mass is conserved over time.
Then,  the conservation of mass is ensured by the continuity equation:
\begin{equation}
\frac{\partial{\rho}} {\partial{t}}+ \nabla.({\rho}{v})=0. \label{eqn:2}
\end{equation}
Then, from Eq. (\ref{eqn:2}) we have
\begin{equation}
\nabla.({\rho}{v})= 0, \label{eqn:3}
\end{equation}
due to steady-state and 
where $\rho$ is the density and $v=(v_{R},v_{\phi},v_{z} )$ of the systems.
Here, from the axisymmetric disc we have
\begin{equation}
\frac{1}{R}\frac{\partial}{\partial{R}}(R\Sigma v_{R})=0, \label{eqn:4}
\end{equation}
where $\Sigma $ is the surface density and for a steady state, Eq. (\ref{eqn:4}) yields an accretion rate as:
\begin{equation} 
\dot{M} =-2\pi R\Sigma v_{R}= constant, \label{eqn:5}
\end{equation}

\subsection {Angular momentum conservation}
Assuming a steady-state, the Navier- Stoke's equation can be expressed as:
\begin{equation}
\rho(v.\nabla)v=-\nabla p +\rho\nabla\phi + J\times B +\nabla.(\rho v(\nabla v-\frac{2}{3}(\nabla. v))), \label{eqn:6}
\end{equation}
where \rm{ p} is pressure, $v$  kinematic viscosity, $\phi$ the gravitational potential
$J=\frac{1}{\mu_0}(\nabla\times B)= (J_R,J_\phi,J_z)$ the current density and $B=(B_R, B_\phi,B_z)$ the magnetic field.
Here, we only consider the azimuthal component of the Navier- Stoke's equation, which is given by:
\begin{equation}
\Sigma\left(\frac{\partial v_{\phi}}{\partial t} +\frac{v_R}{R}\frac{\partial}{\partial R}(R B_{\phi})\right)=
 \frac{B_R}{\mu_0}\frac{1}{R}\frac{\partial}{\partial R}(R B_{\phi})
 + \frac{B_z}{\mu_o}\frac{\partial B_{\phi}}{\partial z} +
 \frac{1}{R^2}\frac{\partial}{\partial R}\left(R^3\Sigma v\frac{\partial}
 {\partial R}(\frac{v_{\phi}}{R}\right), \label{eqn:7}
\end{equation}
Here, we neglect $\frac{B_R}{R}
\frac{\partial}{\partial R}(R B_{\phi})$ and for steady-state disc $\frac{\partial}{\partial t}=0$. 
By integrating Eq. (\ref{eqn:7})  and multiplying both sides by R \citep{Tessema+2010structure},
we get the angular momentum conservation:
\begin{equation}
\Sigma\left(v_R\frac{dl}{dR}\right) =\left[\frac {B_zB_\phi}{\mu_0}\right]^H_{-H} R+
\frac{1}{R}\frac{d}{dR}\left[R^3v\Sigma\frac{d}{dR}(\frac{l}{R^2})\right], \label{eqn:8}
\end{equation}
where $l=Rv_\phi\propto R^{1/2}$ is the specific angular momentum. 
Then, the magnetic field of the NS in the \citep{Wang+1995torque} 
is given by:
\begin{equation}
B_{z}=-\frac{\mu}{R^3}. \label{eqn:9}
\end{equation}
where $\mu$ is the magnetic dipole moment.\\
Here, from Eq. (\ref{eqn:8}) we have two sources of magnetic fields, $B_{\phi}$,  such as shear magnetic field, 
$B_{\phi}, shear$ and dynamo generated magnetic field, $B_{\phi}, _{dyn}$ \citep{Balbus+1998instability}. 
As it was proposed by \citep{Wang+1995torque} the magnetosphere is nearly force free, and reconnection takes place outside the disc.
The ratio of vertical and azimuthal field strengths is related to the shear between the disc and the magnetic field.
Then, this ratio can be expressed in the form of \cite{livio+1992dwarf}:

\begin{equation}
\frac{B_{\phi, shear}}{B_z}\sim -\gamma(\frac{\Omega_k - \Omega_s}{\Omega_k}), \label{eqn:10}
\end{equation}
where $\Omega_k$ and $\Omega_s$  represents the Keplerian angular velocity at the inner radius of the disc and
the angular velocity of the star, respectively. The subscript k- denotes the keplarian.
By rearranging Eq. (\ref{eqn:10}) we obtain:  
\begin{equation}
B_{\phi}, _{shear}=_-\gamma B_z\left(\frac{\Omega_k- \Omega_s}{\Omega_k}\right), \label{eqn:11}
\end{equation}
where $\gamma$ is a dimensionless parameter of  a system \citep{Ghosh+1979accretion}.
The dynamo  magnetic field, $B_{dyn}$, generated by magnetohydrodynamical turbulence in accretion disc through the dynamo 
action \citep{Balbus+1998instability}, which is given by:
\begin{equation}
B_{\phi}, _{dyn} =\epsilon (\alpha_{ss}\mu_{0}\gamma_{dyn} P(r))^1/2, \label{eqn:12}
\end{equation}
where the subscript dyn - is stands for the dynamo generated magnetic field and $ P(r)$ is the radiation pressure.
From Eq. (\ref{eqn:12}) $\gamma_{dyn}$ is the order of 10 that is given by \cite{Brandenburg+1995dynamo}
and $\epsilon$ is a dynamo parameter that describes the direction of the magnetic field 
in the range of $-1\leq \epsilon \leq 1$. Then, substituting Eqs. (\ref{eqn:9}),  (\ref{eqn:11}) and 
(\ref{eqn:12}) into Eq. (\ref{eqn:8}) and from \cite{Tessema+2010structure}  we have obtained:
\begin{equation}
\Sigma\left(v_R \frac{dl}{dR}\right)= 2\epsilon\frac{(\mu R^{-3})}{\mu_0}(\alpha_{ss}\mu_0\gamma_{dyn} P(r))^1/2R 
-2\gamma\frac{(\mu R^{-3})^2}{\mu_0}\left(\frac{\Omega_k-\Omega_s}{\Omega_k}\right)R +
\frac{1}{R}\frac{d}{dR}\left(R^3 v\Sigma\frac{d}{dR}(\frac{l}{R^2})\right). \label{eqn:13}
\end{equation} 
 Eq. (\ref{eqn:13}) shows an ordinary differential equation for an accretion disc of the angular momentum  consevation.
 
 \subsection{Hydrostatic vertical balance}
We now consider the structure of the disc in the vertical z-direction. Hence, the angular momentum conservation is reduced
to hydrostatic equilibrium condition if the net flow of gas along the vertical direction is zero. Then, the hydrostatic 
equilibrium equation is given by
\begin{equation}
\frac{1}{\rho}\frac{\partial P}{\partial z}=\frac{\partial}{\partial z}\left(\frac{G M}{(R^2+ z^2)}\right)^{1/2}, \label{eqn:14}
\end{equation}
in the limit $z\ll R$ and neglecting the self-gravity of the disc, Eq. (\ref{eqn:14}) becomes:
\begin{equation}
\frac{1}{\rho}\frac{\partial P}{\partial z}= -\frac{G Mz}{R^3}. \label{eqn:15}
\end{equation}
where $\rm{G}$, $\rm{M}$  are the universal gravitational constant and  a mass of the accreting star
, respectively.

As a consequence, from Eq. (\ref{eqn:1}) and the approximation of vertical pressure gradient we express  
$\frac{\partial P}{\partial z}\sim \frac{p}{H}$ and  $z\sim H$. Then, Eq. (\ref{eqn:15}) yields:
\begin{equation}
\frac{P}{\rho}= c_s^2, \label{eqn:16}
\end{equation}
Thus, from Eq. (\ref{eqn:15}) and Eq. (\ref{eqn:16}) we find the $H$ as:
\begin{equation}
H\cong c_sR\left(\frac { R}{GM}\right)^{1/2}. \label{eqn:17} 
\end{equation}
For a thin accretion disc, the local Kepler velocity should be highly supersonic.
In general, we can define a central disc density approximately by
\begin{equation}
\rho=\frac{\Sigma}{H}~~and~~
H= c_s\left(\frac{R}{v_{\phi}}\right), \label{eqn:18}
\end{equation}
where $v_{\phi}$ is given by
\begin{equation}
v_{\phi}=\sqrt{\left(\frac{GM}{R}\right)}. \label{eqn:19}
\end{equation}
The speed of sound can be expressed by using Eq. (\ref{eqn:18}) and (\ref{eqn:19}) as:
\begin{equation}
c_s=\frac{H}{R}\left(\frac{GM}{R}\right)^{1/2}, \label{eqn:20}
\end{equation}
As it was proposed by \cite{Tessema+2010structure} we can write the gas and radiation pressure as:
\begin{equation}
P=\frac{\rho K_B T_c}{\overline{\mu} m_p}+ \frac{4\sigma}{3c}T^4_c, \label{eqn:21}
\end{equation}
where $ \sigma$ is the Stefan -Boltzmann constant, $m_p$ is the mass of a proton , $T^4_c$ central temperature,
the subscript c- denotes values in the central plane, 
c is the speed of light, $\overline \mu$ is the mean molecular weight for ionized gas and $K_B$ is the Boltzmann's 
constant.
Then, we can write the pressure for hydrostatic equilibrium using Eqs. (\ref{eqn:15}) and (\ref{eqn:18}) as:
\begin{equation}
P=\Sigma\left(\frac{ HGM}{R^3}\right).  \label{eqn:22}
\end{equation}

Here, for a newtonian an accretion disc,  the $f_{R\phi}$ component of the viscous stress tensor is given by:
\begin{equation}
f_{R\phi}= \frac{-3}{2} \rho v\Omega =\alpha_{ss} P(r), \label{eqn:23}
\end{equation}
where
\begin{equation}
\Omega= \left(\frac{G M}{R^3}\right)^\frac{1}{2}. \label{eqn:24}
\end{equation}
Substituting Eq. (\ref{eqn:24}) into Eq. (\ref{eqn:23}) we obtain:
\begin{equation}
f_{R\phi}= \frac{-3}{2} \rho v\left(\frac{G M}{R^3}\right)^\frac{1}{2}= \alpha_{ss}P(r), \label{eqn:25}
\end{equation}
where, $\rho=\frac{\Sigma}{2 H}$.
Then, the viscous stress tensor,$f_{R\phi}$, can be expressed by:
\begin{equation}
f_{R\phi}=\frac{3\Sigma v}{4H}\left(\frac{GM}{R^3}\right)^{1/2}=\alpha_{ss}P(r). \label{eqn:26}
\end{equation}
From Eqs. (\ref{eqn:16}), (\ref{eqn:22}) and (\ref{eqn:26}) we obtained the gas density of the NS as:
\begin{equation}
\rho=\frac{3v\Sigma}{4\alpha_{ss} H^3}\left(\frac{GM}{R^3}\right)^{-1/2}. \label{eqn:27}
\end{equation}
Also, we can express the scale height of the disc in terms of the total pressure as:
\begin{equation}
H= \left(\frac{\rho k_B T_c R^3}{m_p\overline\mu GM}+\frac{4\sigma T^4_c R^3}{3c\rho GM}\right)^{1/2}. \label{eqn:28}
\end{equation}
The local viscous dissipation is determined by radiative losses when the matter flow through an optical disc is low.
Then, we have the $T_{c}$, $v$, $\Sigma$, $M$ and $R$ relations: 
\begin{equation}
\frac{4\sigma}{3\tau}T^4_c=\frac{9}{8}v\Sigma\frac{GM}{R^3}. \label{eqn:29}
\end{equation}
Here, the optical depth of the disc,$\tau$, is given by
\begin{equation}
\tau=\int^H_0{k_R\rho dz}=\rho H k_R, \label{eqn:30}
\end{equation} 
where $K_{R} =k_{es}+k_{ff}$, in the inner region of the disc the temperature is high the approximation of
$k_R\approx k_{es}$ is valid becuase this region is dominanted by electron scattering opacity.
Then, from Eqs. (\ref{eqn:29}) and (\ref{eqn:30}) we obtain the centeral temperature
\begin{equation}
 T^4_c=\frac{27}{32\sigma}v\Sigma^2 k_R\frac{GM}{R^3}. \label{eqn:31}
\end{equation}
As it was investigated by \citep{Tessema+2010structure}; \cite{Frank+2002accretion}; 
\cite{Shapiro+1983implications}:
\begin{equation}
R_A=\left(\frac{2\pi^2\mu^4}{GM\dot{M^2}\mu_0^2}\right)^\frac{1}{7}\simeq 
1.4\times10^4\dot{M_{14}}^\frac{-2}{7} M_1^\frac{-1}{7}\mu^\frac{4}{7}_{15}m. \label{eqn:32}
\end{equation}
\begin{equation}
R_{co}=\left(\frac{GM P^2_{spin}}{4\pi^2}\right)^\frac{1}{3}\simeq 1.5\times 10^6 
P^\frac{2}{3}_{spin} M_1^\frac{1}{3} m, \label{eqn:33}
\end{equation}
where $ P_{spin}=\frac{2\pi}{\Omega_s}$ and $M_1 =\frac{M}{M_{\odot}}$.
Let us to  introduce a parameter $y=\Sigma v$ in order to solve an ordinary differential equation 
for an accretion disc.   
Then, from Eqs. (\ref{eqn:13}), (\ref{eqn:22}),  (\ref{eqn:27}),  (\ref{eqn:32}) and  (\ref{eqn:33}) we have obtained:
\begin{equation}
y^{'}= \frac{\dot{M}}{6\pi r} - \frac{y}{2r} - \epsilon D_1(Gm)^{\frac{-1}{4}}R_A^{\frac{-3}{2}} -D_2 R_A^{\frac{-9}{2}}\left[1 
-\left(\frac{R_A}{R_{co}}\right)^{\frac{3}{2}}\right]. \label{eqn:34}
\end{equation}
where, 
\begin{equation}
D_1=\sqrt{\left(\frac{4\mu^2\gamma_{dyn}y}{3\mu_0 HR_A^{3/2}}\right)}~~~ and ~~~D_2=
\frac{4\mu^2\gamma}{3\mu_0 (Gm)^{1/2}}, \label{eqn:35}
\end{equation}
which is a differential equation of y for accretion disc around accreting millisecond X-ray pulsars.
At large radii the solution of Eq. (\ref{eqn:34}) approaches the Shakura- Sunyaev solution, which giving us the boundary
condition $y\rightarrow \Lambda\dot{M}$ as $R\rightarrow \infty$.
Here, we need to transform Eq. (\ref{eqn:34}) by introducing dimensionless quantities $\Lambda $ and r, so that
\begin{equation}
y=\Lambda \dot{M} \label{eqn:36}
\end{equation}
where $\Lambda$ is a dimensionless parameter for accretion disc and
\begin{equation}
  R= rR_A   \label{eqn:37}
\end{equation}
where $r$ is a dimensionless radial coordinate and $R_A$ is the Alfven radius, which is a characteristic radius at
which magnetic stresses dominate the flow in the accretion disc.
As noted by \cite{Elsner+1977accretion} we have $\omega_s$ as: 
\begin{equation}
\omega_s=\left(\frac{R_A}{R_c}\right)^\frac{3}{2}=
0.36M^\frac{-5}{7}_1\dot M^\frac{-3}{7}_{14}\mu^\frac{6}{7}_{15}\left(\frac{P_{spin}}{4.8ms}\right)^{-1}, \label{eqn:38}
\end{equation}
Finally, using Eq. (\ref{eqn:36}), Eq. (\ref{eqn:37}) and Eq. (\ref{eqn:38}) we get the differential equation of an accretion disc from  Eq. (\ref{eqn:34}) 
which is given by:
\begin{equation}
\Lambda^{'}=\frac{1}{6\pi r}- \frac{\Lambda}{2 r}- \epsilon D_3(GM)^\frac{-1}{4}R_A^\frac{-5}{4} r^\frac{-9}{4}-
D_4 r^\frac{-9}{2}\left(1- \omega_s r^\frac{3}{2}\right). \label{eqn:39}
\end{equation}
where $D_3=\sqrt{\left(\frac{4\mu^2\gamma_{dyn}\Lambda}{3\mu_0 H\dot{M}}\right)}$ and $ D_4=
\frac{4\mu^2\gamma}{3\mu_0(Gm)^\frac{1}{2}\dot{M}} R_A^\frac{-7}{2}$.
This equation is the new analytical  solution for an accretion disc around accreting millisecond X-ray pulsars.

 \subsection{The structure of the disc}
Here, to analyze the dynamics of an accretion disc, we emphasis on the inner region of the disc, in which
the radiation pressure is much higher than the gas pressure and the accretion rate is large.
In this region, Compton scattering occurs more frequently than free-free absorption.
To solve Eq. (\ref{eqn:39}) numerically we have to determine scale height in the inner region of the disc, which is
given by:
\begin{equation}
H=\frac{9}{8c}k_{es} (\dot M\Lambda). \label{eqn:40}
\end{equation}
Then, the shear magnetic field is given by:
\begin{equation}
B_{\phi}, _{shear}= -4\times10^3\gamma M^\frac{3}{7}_1 
\dot M^\frac{6}{7}_{14}\mu^\frac{-5}{7}_{15}\left(1-\omega_s r^\frac{3}{2}\right) r^{-3} T, \label{eqn:41}
\end{equation}
In the inner region of the  disc the radiation pressure is larger than the gas pressure, we have that:
\begin{equation}
\Sigma=9.57\times10^1\alpha_{ss}^-1M_1^{-5/7}\dot{M}^{-10/7}_{14}\mu^{6/7}_{15}\Lambda(r)^{-1}r^{3/2} kgm^{-2}, 
\label{eqn:42}
\end{equation}
\begin{equation}
\rho_c=3.18\times10^{-2}\alpha_{ss}^-1M^{-5/7}_1\dot{M}^{-17/7}_{14}\mu^{6/7}_{15}\Lambda(r)^{-2}r^{3/2}kgm^{-3}, 
\label{eqn:43}
\end{equation}
\begin{equation}
v_R=1.18\times10^7\alpha_{ss}M^{6/7}_1\dot{M}^{19/7}_{14}\mu^{-10/7}_{15}\Lambda(r)r^{-5/2}ms^{-1},\label{eqn:44}
\end{equation}\begin{equation}
T_c=1.86\times10^6\alpha_{ss}^{-1/4}M^{5/28}_1\dot{M}^{3/28}_{14}\mu^{-3/4}_{15}r^{-3/8} \rm{k}, \label{eqn:45}
\end{equation}
\begin{equation}
v=1.1\times10^{12}\alpha_{ss}M^{5/7}_1\dot{M}^{17/7}_{14}\mu^{-6/7}_{15}\Lambda(r)^2r^{-3/2} m^2s^{-1}, \label{eqn:46}
\end{equation}
\begin{equation}
\tau_{es}=1.86\alpha_{ss}^{-1}M^{-5/7}_1\dot{M}^{-10/7}_{14}\mu^{6/7}_{15}\Lambda(r)^{-1}r^{3/2}, \label{eqn:47}
\end{equation}
The transition radius in the inner region of the disc is estimated by approximating 
$\Lambda$ = $1/3\pi$.
\begin{equation}
r_{IM}=12.5\overline{\mu}^{8/21}\alpha^{2/21}_{ss}M^{10/21}_1\dot{M}^{22/21}_{14}\mu^{-4/7}_{15}, \label{eqn:48}
\end{equation}
Here, the accretion disc outside of Alfven radius is overpowered by radiation pressure only if
\begin{equation}
\mu_{15}<82.56\overline{\mu}^{2/3}\alpha^{-1/16}M^{5/6}_1\dot{M}^{11/6}_{14}, \label{eqn:49} 
\end{equation}
This circumstance is not satisfied for ordinary X-ray pulsar with a magnetic dipole moment
of $\sim10^{20} \rm{Tm^3}$ \cite{White+1988radius}, though it can be satisfied for AMXPs.

As we have incorporated so far about the equations of an accretion disc around accreting millisecond X-ray pulsars, then we
applied some parameters and investigate Eq. (\ref{eqn:39}) in the inner region of the disc.
The inner region in which the radiation pressure is overpowered and electron scattering is the most important source of
opacity \cite{Shakura+1973black}.  As a result, in the innermost regions, the emitted 
spectrum of the disc cannot be approximated by a black-body spectrum.
whereas if the accretion rates are high, radiation pressure towards the inner
edge of the accretion disc exceeds the thermal pressure.
Using the appropriate selection of the magnetic field and accretion rate, then the inner region solution is found below.

\section{Result and Discussion}
\subsection{Global Solutions}
\begin{figure}
	\centering
	\includegraphics[width=0.8\textwidth, angle=0]{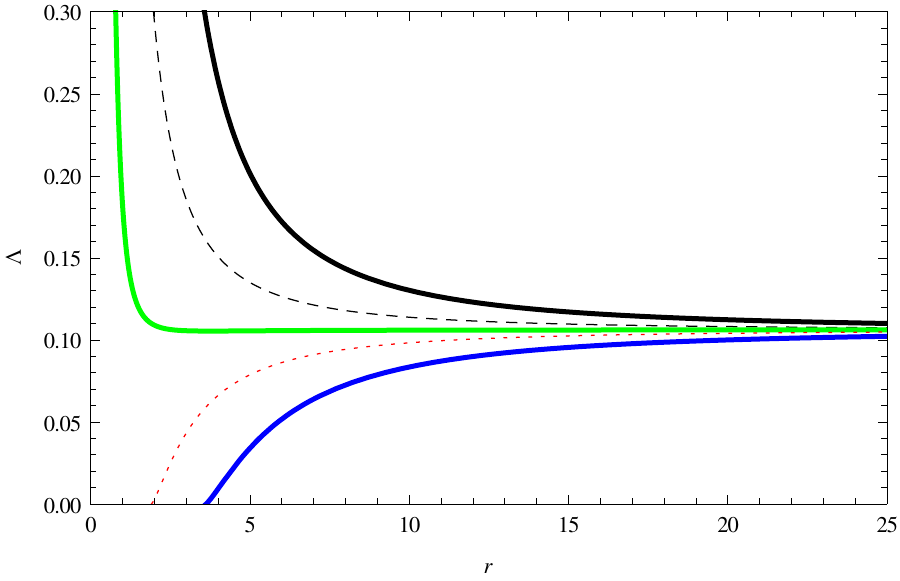}
	\caption{Result of $\Lambda$  as a function of $r$ for the AMXPs  with accretion rate $\rm{\dot{M}}=1.5\times 10^{14}\rm{kgs^{-1}}$ and the different dynamo parameters are shown with $\epsilon= -0.45$ solid blue line, $\epsilon=- 0.15$ red dotted line, $\epsilon =0$ solid green line, $\epsilon= 0.15$ black dotted line, and $\epsilon= 0.45$ solid black line.}
        \label{Fig1}
       \end{figure}

Here, as it was studied by \cite{Tessema+2011thin} we integrate Eq. (\ref{eqn:39}) for the inner region inwards from 
very small radius $\sim 12.5$ and $\Lambda=1/3\pi$. 
The dimensionless parameters $\gamma$, $\gamma_{dyn}$  and $\alpha_{ss}$ are   $1$, $10$ and $10^{-2}$, respectively.
Hence, the disc is overpowered by radiation pressure and electron scattering 
which $R_A<R_{IM}$ by increasing accretion rate. In this region, it is possible to use the accretion rate up 
to Eddington limit, so that we take $\rm{\dot{M}}=1.5\times10^{14} \rm{kgs^{-1}}$ for our calculation and we use $\Lambda=1/3\pi$  
for the analytical solution of the disc, 
 and solve Eq. (\ref{eqn:39}) for the inner region of the disc starting from $\rm{r_{IM}}=12.5$.
Our solution for the inner disc region of AMXPs with  different dynamo parameters such as $\epsilon =0.45,0.15, 0,-0.15, -0.45$ 
are shown in Fig.~\ref{Fig1}. This figure shows the variations of $\Lambda$ as a function of $r$ for all $\epsilon = 0.45,0.15$ and  $0$ 
all solutions are case v inner boundaries except $-0.15,-0.45$ \cite{Tessema+2011thin}.\\
 
\begin{figure}
	\centering
	\includegraphics[width=0.8\textwidth angle=0]{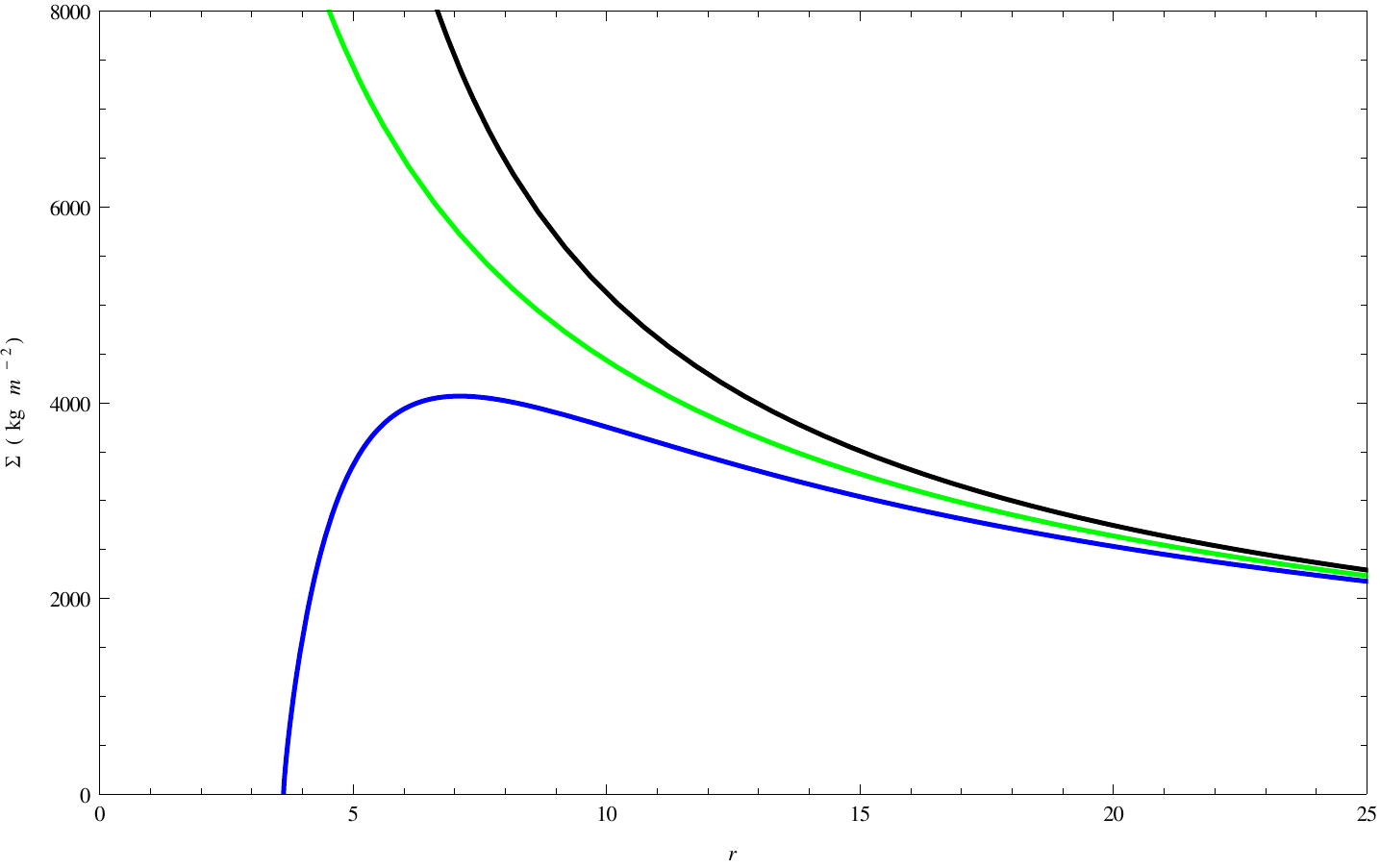}
	\caption{Result of  $\Sigma$ as a function of $r$ for AMXPs with accretion rate $\rm{\dot{M}}=1.5\times 10^{14}\rm{kgs^{-1}}$ and the  different dynamo parameters are shown with $\epsilon= -0.45$ solid blue line,$\epsilon = 0$ solid green line,  and $\epsilon = 0.45$ solid black line.}
        \label{Fig2}
        \end{figure}

In Fig.~\ref{Fig2} the $\Sigma$ is purely a decreasing function of $r$ for $\epsilon= 0, 0.45$ and increasing for $\epsilon=-0.45$.
 
\begin{figure}
	\centering
	\includegraphics[width=0.8\textwidth, angle=0]{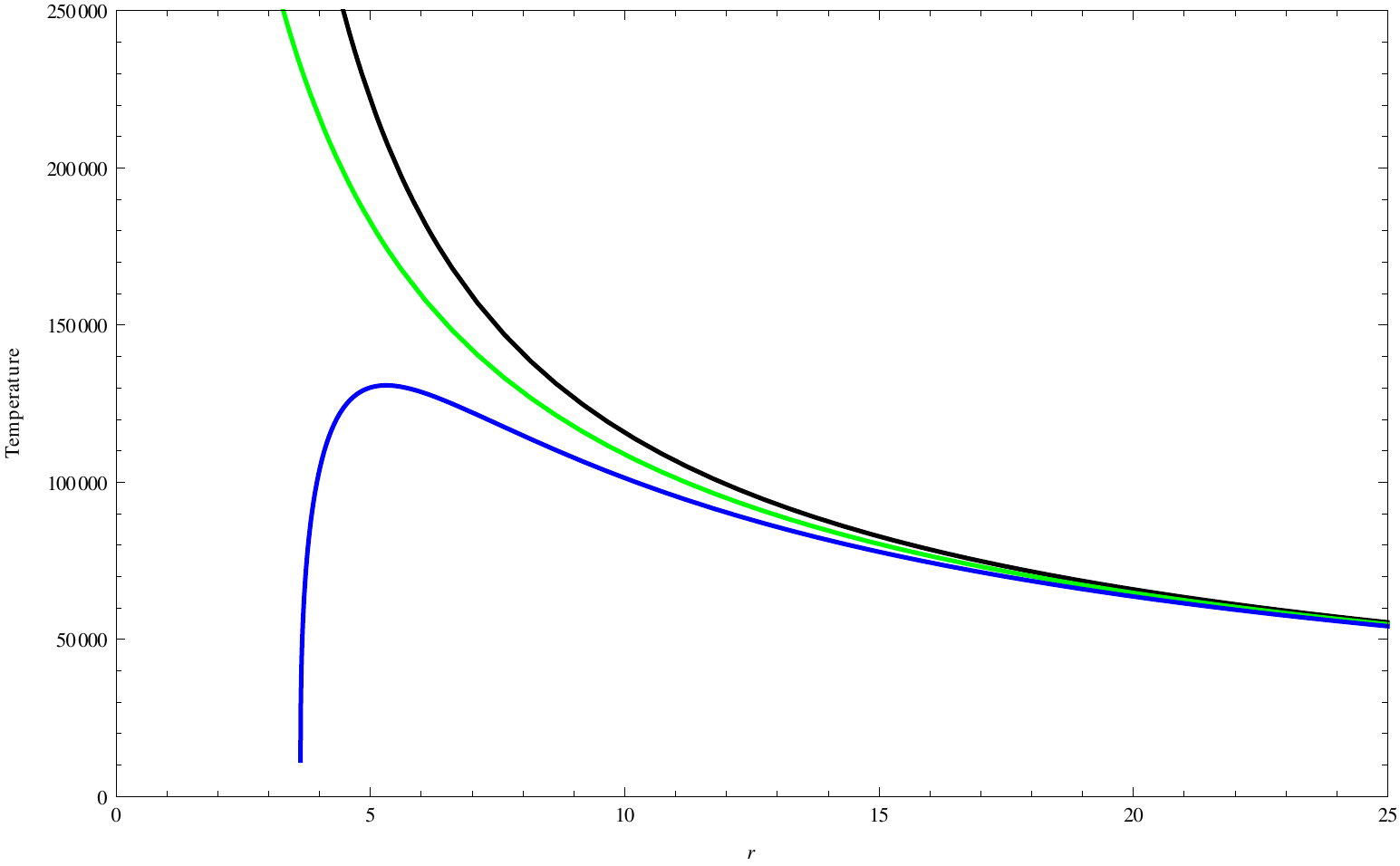}
	\caption{Result of $\rm{T_{c}}$  as a function of $r$ for AMXPs with accretion rate $\rm{\dot{M}}=1.5\times 10^{14}\rm{kgs^{-1}}$ and the  different dynamo parameters are shown with $\epsilon= -0.45$ solid blue line, $\epsilon = 0$ solid green line, and $\epsilon = 0.45$ solid black line. }
	\label{Fig3}
	\end{figure}

In Fig.~\ref{Fig3} the high surface density results in a hot flow. Here,  the mid-plan temperature as a function of the radius in the inner region of the disc does not depend on $\Lambda$ so that as the radius decreases the temperature increases.

\begin{figure}
	\centering
	\includegraphics[width=0.8\textwidth, angle=0]{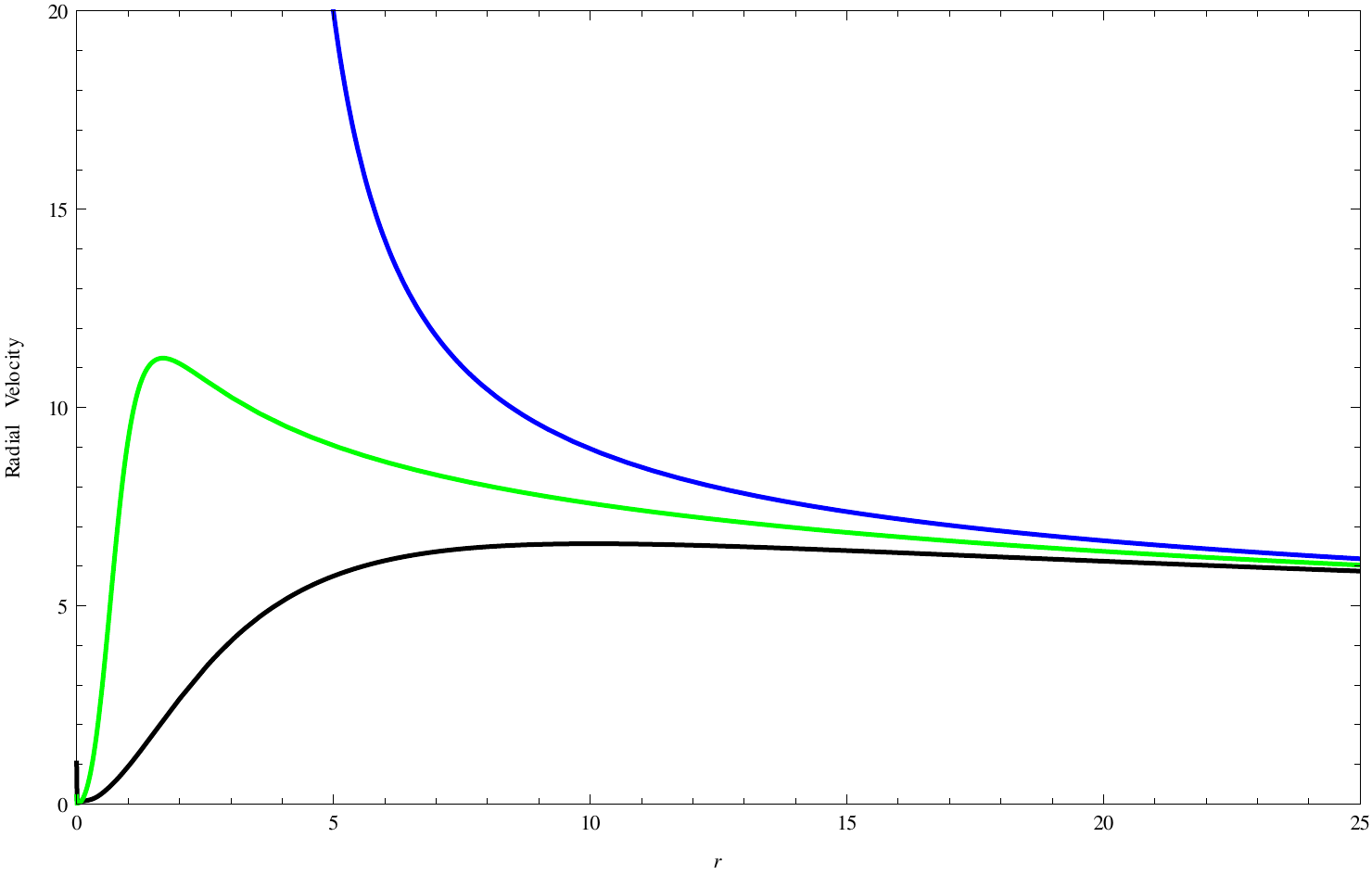}
	\caption{Result of  $V_R$  as a function of r for AMXPs with accretion rate
$\rm{\dot{M}}=1.5\times 10^{14}\rm{kgs^{-1}}$ and the different dynamo parameters are shown with  $\epsilon= -0.45$  solid blue line,
$\epsilon = 0$ solid green line, and $\epsilon= 0.45$ solid black  line}
	\label{Fig4}
	\end{figure}

In Fig.~\ref{Fig4} we investigated that the high surface density which is corresponding to a decreases 
in radial velocity and this radial velocity is dependent on $\Lambda$. On this  figure, the inner edge of the accretion 
disc approaches to the surface of the star the radial velocity either goes to zero or infinite.
Here, the radial velocity decreases as the radius increases.\\

\subsection{Accretion torques}
\begin{figure}
	\centering
	\includegraphics[width=0.8\textwidth, angle=0]{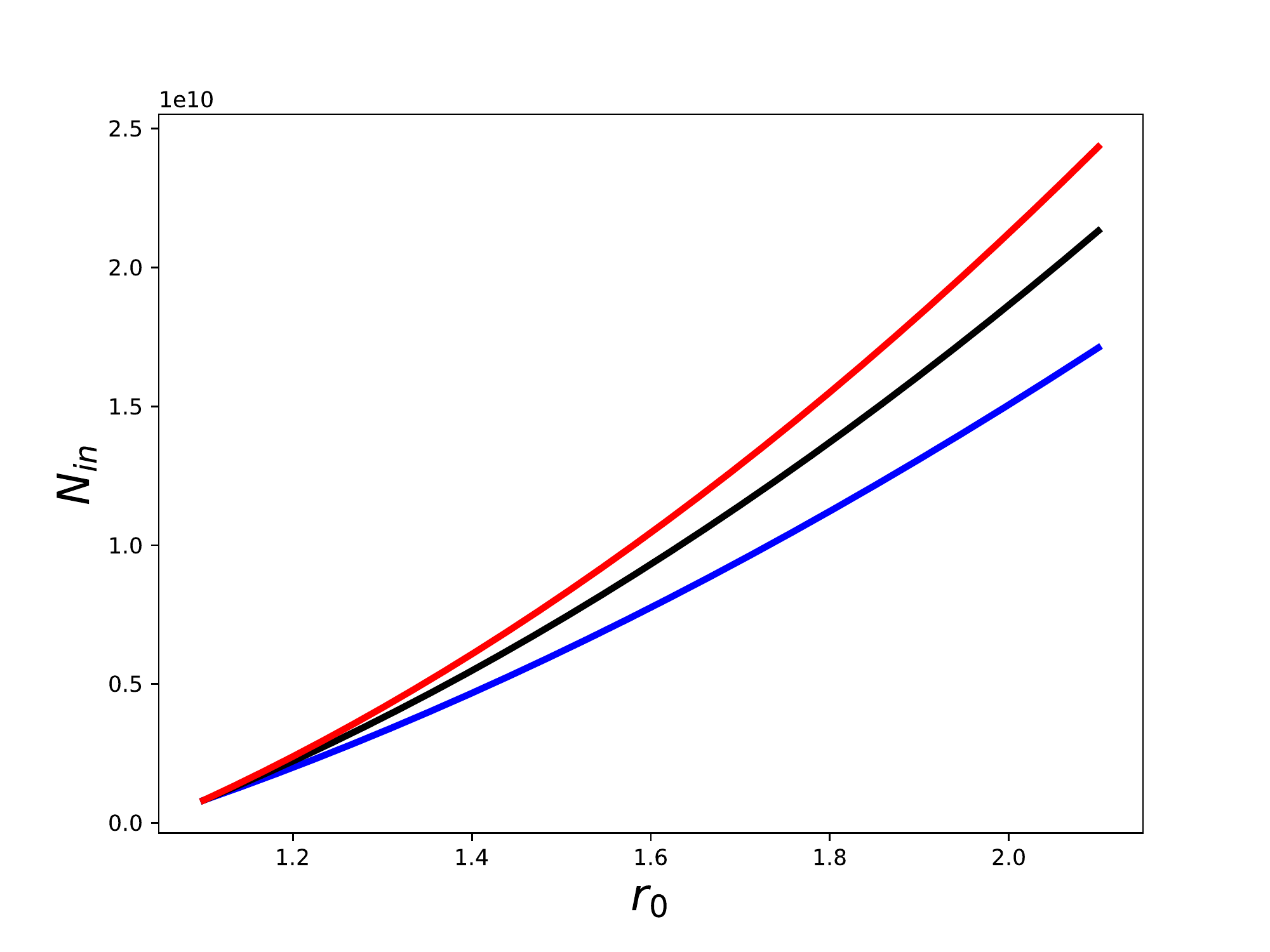}
	\caption{shows the material torque on  AMXPs with different accretion rates by varying  the inner region  radius, $ r_{in}$, of the disc}
	\label{Fig5}
	\end{figure}

The torques on a neutron star range from material to magnetic; It is obtained from Eq. (\ref{eqn:13}) by multiplying
$2\pi R$ and then integrating from the inner radius of the disc, $\rm{R_{in}}$, to  the outer edge
of the disc, $\rm{R_{out}}$, see, e.~g., \cite{kluzniak+2007magnetically}; \cite{Tessema+2010structure};  
\cite{shi+2015super}.  
\begin{equation}
\dot{M}\sqrt{GMR_{in}}-\dot{M}\sqrt{GMR_{out}} =-\int_{R_{in}}^{R_{out}}\left[\frac{4\pi 
(\mu R^{-3})}{\mu_0}(B_{\phi,dyn} +
B_{\phi,shear})\right] R^2dR -\left[3\pi y(GMR)^{1/2}\right]_{R_{in}}^{R_{out}}, \label{eqn:50}
\end{equation}
Note that the two expressions on the  LHS of Eq. (\ref{eqn:50}) show the rate at which angular momentum is 
transported past 
the inner and outer edges of the accretion disc, while the RHS shows the implications of magnetic and viscous torques to 
the angular momentum balance.
In this case, the material,  
magnetic  and viscous torque can be expressed in Eqs. (\ref{eqn:51}), (\ref{eqn:52}), (\ref{eqn:53}), 
(\ref{eqn:54}), (\ref{eqn:55}) and (\ref{eqn:56}):
Thus, the material torque of the inner edge of the disc on the neutron star is given by
\begin{equation}
N_{in}=\dot{M}(GMR_{in})^{1/2}=1.4\times 10^{26}\mu_{15}^{2/7}M_1^{3/7}\dot{M}_{14}^{6/7}r_{in}^{1/2}, \label{eqn:51}
\end{equation}
Here, in Fig.~\ref{Fig5} we investigate the inner accretion torque on the disc in the inner region of the disc.
This material torque increases as the accretion rate and the inner radius increases.
 
The magnetic torque is the result of the coupling between the vertical magnetic field of the
star and the toroidal magnetic field in the disc.
Hence, the torque acting on the lower surface of the disc can be written  \citep{Ghosh+1979accretion} as:
\begin{equation}
N_{mag}= -4\pi\int_{R_{in}}^{R_{out}} \frac{-(\mu R^{-3}) \left(B_{\phi,dyn}+B_{\phi,shear}\right) 
 R^2dR}{\mu_0}, \label{eqn:52}
\end{equation}
This magnetic torque is separated into the shear and dynamo generated magnetic torque. 
Then, the shear magnetic torque, $N_{mag, shear}$, is given by:  
\begin{equation}
N_{mag, shear}=\int_{R_{in}}^{R_{out}} 4\pi\frac{-(\mu R^{-3}) B_{\phi,shear}}{\mu_0}R^2dR 
\approx 4\times10^{26}\gamma\mu_{15}^{2/7}M_1^{3/7}\dot{M}_{14}^{6/7}\int_{r_0}^{\infty}[r^{-4}(1 - 
\omega_sr^{3/2})]dr, \label{eqn:53}
\end{equation}
and the dynamo generated  magnetic torque on the neutron star is:
\begin{equation}
N_{mag, dyn}=-\int_{R_{in}}^{R_{out}}4\pi\frac {-(\mu R^{-3}) B_{\phi,dyn}}{\mu_0}R^2dR. \label{eqn:54}
\end{equation}

Here, the dynamo generated magnetic torque in the inner region of the disc is give by:
\begin{equation}
N_{dyn,inner}= 7\times10^{26}\epsilon\gamma_{dyn}^{1/2}\mu_{15}^{4/7}M_1^{5/14}\dot{M}_{14}^{3/14}\int_{inner}r^{-7/4}dr. 
\label{eqn:55}
\end{equation}
On the other hand, the viscous torque in the inner region of the disc is given by:

\begin{equation}
N_{vis}= -3\pi yR_{in}(GMR_{in})^{1/2}= -1.3\times10^{27}\mu_{15}^{2/7}M_1^{3/7}\dot{M}_{14}^{6/7}
\Lambda(r_0)r_{in}^{1/2}. \label{eqn:56}
\end{equation}
Moreover, as it was investigated  by \cite{Tessema+2011thin} the standard accretion disc solution has a case D 
inner region boundary when the 
viscous torque is neglected in accretion disc theory.
Hence, the angular momentum is transported from the neutron star to the disc when it is in case V inner region boundary.
Except if  $\epsilon=0$ the dynamo magnetic torque importantly greater than the shear magnetic torque, and both are greater for
 $\epsilon=0.15$ than for $\epsilon=0.45$. Because of this effect the central hole of the disc is grows too large when 
 $\epsilon=0.45$. The overpowered torque at $\epsilon =0.45$ is the viscous torque at this region, which is ignored as show in
 Fig.6 below.
	\begin{figure}
	\centering
	\includegraphics[width=0.8\textwidth, angle=0]{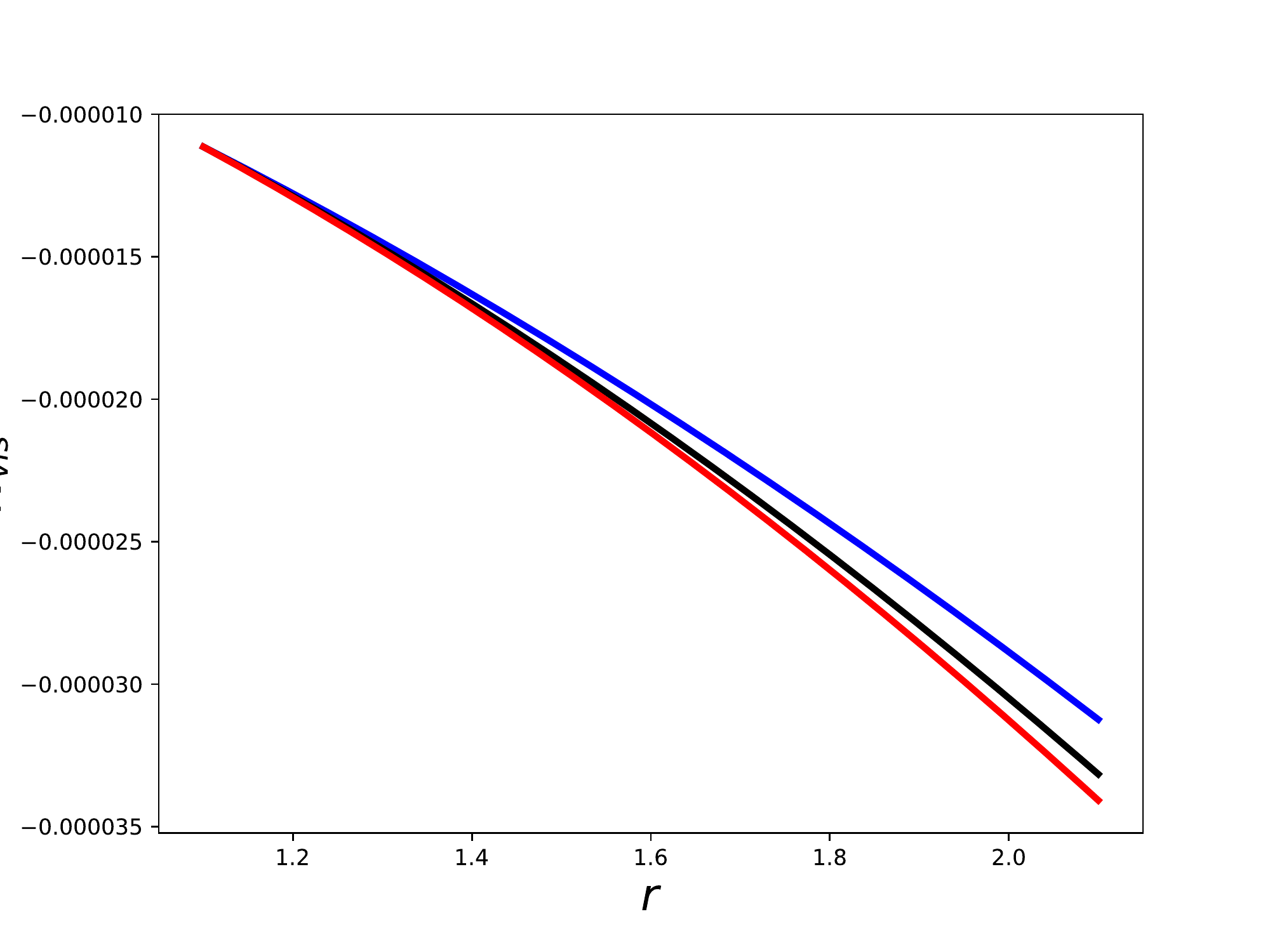}
	\caption{shows the variation of viscous torque as a function of radius for AMXPs with differet masses and radius in the inner 
region of the disc}
	\label{Fig6}
	\end{figure}

\subsection{Comparison with observational Results}
There is a large variation in the accreting rates among the accreting millisecond X-ray pulsars. 
The well studied system IGR J00291+5934 is accreting at a rate of at least $\sim 10^{14}\rm{kgs^{-1}}$ based on 
its x-ray flux \cite{Burderi+2007timing}, while in some other systems, for instance, SAX J1808.4-3658
\cite{Bildsten+2001brown}, the neutron star is accreting at a rate below $10^{12} \rm{kgs^{-1}}$ from
a brown dwarf companion.
There is also a great doubtful in the spin variations that have been reported for the AMXPs. 
For instance, \cite{Burderi+2006order} reported $\dot{v}s$
these spin variations
between $-7.6\times 10^{-14}$ and $4.4\times 10^{-13}\rm{ Hzs^{-1}}$ for $SAX J1808.4-3658$.
But \cite{Hartman+2008long} noted that the measurements of this source are plagued by more variations in
the pulse shape, and put an upper limit of $2.5\times 10^{-14}\rm{Hzs^{-1}}$ on the spin variations and in
found along -term spin down $\dot{v}=-5.6\times10^{-16} \rm{Hzs^{-1}}$.

On the other hand \cite{Burderi+2007timing} reported that $IGR J00291+5934$ is spinning up at
$\sim10^{-12}\rm{Hzs^{-1}}$ during the December 2004 outburst. The more spin variations that have been observed
in some AMXPs depends on the accreting torque, which is given by:
\begin{equation}
N=2\pi\dot{v}I, \label{eqn:57}
\end{equation}
where $I$ is in 
$\rm{Kg m^2}$, and $\dot v$ is 
in  $\rm{Hzs^{-1}}$.

\section{Conclusion}
In this paper, we have studied the interaction between the accreting millisecond X-ray pulsars 
and the inner region of the disc, which is supported by the dynamo generated magnetic field.
Hence, we found that the fundamental equations of an accretion disc around accreting millisecond X-ray pulsars gives 
the more stable system than the previous study.
We have made an effort to find an analytical solution by using a numerical method for an accretion disc around AMXPs 
in the inner region of the disc, in which the accretion rate is high and the disc overpowered by radiation pressure 
and electron scattering region.
Here, the  analytical solution of Eq. (\ref{eqn:39}) at higher accretion rate  in the inner region of 
the accretion disc is greater than the radius of the neutron star for different values of dynamo parameters, $\epsilon$ and
we observed the behavior of these solutions in the inner region  in  Fig.~\ref{Fig1}.  
We have formulated the numerical method for structure equation in the inner region of the disc and the highest accretion rate
is sufficient to make the innermost region of the accretion disc to be overpowered by radiation pressure and electron scattering.
We have observed that the relationship between surface density and radius in Fig.~\ref{Fig2}. Then, on this figure, the surface
density decreases with the radius.
The viscous torque in the inner region of the disc is ignored,  which is shown in Fig.~\ref{Fig6}. Hence, the viscous torque on
AMXPs decreases for different masses and radius of the accretion disc. 
 The accretion torque is important in explaining the observed variations in the spin frequency of AMXPs 
like IGR J00291+5934.
Thus, we have found that the spin derivatives for accretion rate $1.5\times10^{14}\rm{kg/s}$ in this model are in agreement
with RXTE observed data in accreting millisecond X-ray pulsars are consistently explained by this model.
\begin{acknowledgements}
We thank  Ethiopian Space Science and Technology Institute, Entoto Observatory and Research Center 
and Astronomy and Astrophysics Research and Development Department for supporting this research. This research has made of 
Astronomical Data system.
\end{acknowledgements}

\label{lastpage}

\end{document}